# An ITO – Graphene hybrid integrated absorption modulator on Si-photonics for neuromorphic nonlinear activation

Rubab Amin,[1] Jonathan K. George,[1] Hao Wang,[1] Rishi Maiti,[1] Zhizhen Ma,[1] Hamed Dalir,[1] Jacob Khurgin,[2] and Volker J. Sorger[1,a)]

[1]*Department of Electrical and Computer Engineering, George Washington University, 800 22nd St. N.W., Washington, District of Columbia 20052, USA*

[2]*Department of Electrical and Computer Engineering, Johns Hopkins University, Baltimore, Maryland 21218, USA*

**Abstract**

The high demand for machine intelligence of doubling every three months is driving novel hardware solutions beyond charging of electrical wires given a resurrection to application specific integrated circuit (ASIC)-based accelerators. These innovations include photonic-based ASICs (P-ASIC) due to prospects of performing optical linear (and also nonlinear) operations, such as multiply-accumulate for vector matrix multiplications or convolutions, without iterative architectures. Such photonic linear algebra enables picosecond delay when photonic integrated circuits are utilized, via 'on-the-fly' mathematics. However, the neuron's full function includes providing a nonlinear activation function, knowns as thresholding, to enable decision making on inferred data. Many P-ASIC solutions performing this nonlinearity in the electronic domain, which brings challenges in terms of data throughput and delay, thus breaking the optical link and introducing increased system complexity via domain crossings. This work follows the notion of utilizing enhanced light-matter-interactions to provide efficient, compact, and engineerable electro-optic neuron nonlinearity. Here, we introduce and demonstrate a novel electro-optic device to engineer the shape of this optical nonlinearity to resemble a rectifying linear unit (ReLU) - the most-commonly used nonlinear activation function in neural networks. We combine the counter-directional transfer functions from heterostructures made out of two electro-optic materials to design a diode-like nonlinear response of the device. Integrating this nonlinearity into a photonic neural network, we show how the electrostatics of this thresholder's gating junction improves machine learning inference accuracy and the energy efficiency of the neural network.

<div align="center">**THE MANUSCRIPT**</div>

## I. INTRODUCTION

The growing demands of neural network systems create an urgent need for the development of advanced devices to perform complex operations with fast throughput (operations/s) and lower power dissipation (J/operations), and compact footprint leading to high operation density (operations/s/mm$^2$) [1, 2]. Photonic integrated circuit based artificial neurons can pave the way for this specific challenge. One of the most significant benefits of photonics over electronics is that distinct signals can be straightforwardly and efficiently combined due to their wave-nature exploiting atto-Joule efficient electro-optic (EO) modulators [3–5], phase shifters, and combiners, simplifying essential operations such as weighted sum or addition, vector-

---

[a)] Author to whom correspondence should be addressed. Electronic mail: sorger@gwu.edu.

matrix multiplications or convolutions [6]. In a traditional electrical computing system, the transistors can no longer keep up with the demand for computational complexity since lumped-element electronic circuitry limits the processor speed, alongside the metallic wires will further reduce the signaling delay (transmission throughput). Meanwhile, with device scaling, the static power begins to dominate the power consumption in microprocessors due to subthreshold leakage, a weak inversion current across the device and gate leakage, and a tunneling current through the gate oxide insulation leads to more power consumption and heating [7, 8]. The traditional electronic processor's speed can hardly exceed five gigahertz because of the thermal dissipation limit. Another approach to increase the information processing speed is utilizing parallel computing architectures. Still, the electrical channels (wires) are bound by the physical laws, which cannot carry more than one signal at a time.

In contrast, a photonic system can potentially transmit a few orders of magnitude more information in every square unit owing to the massive parallelism achievable in optics. The reason behind that is photonics operates at a higher baud-rate, due to a lower capacitance (only device and not of the circuit). Also, the superposition property of light allows for further parallelism strategies, where each physical channel can employ multiple wavelengths by exploiting wavelength-division multiplexing (WDM) techniques [9, 10], thus transmitting optical signals in different wavelengths at the same time without occupying extra physical space. This combination can easily achieve a high baud-rate system and aggregated throughput. Besides, optical waves can be modulated in different dimensions, such as by altering the phase, amplitude, mode, wavelength, or polarization.

A neural network usually contains three elements: a set of non-linear nodes (neurons), configurable interconnection (network), and information representation (coding scheme). For a particular non-linear model of a neuron consist of a set of inputs which are the outputs from the other neurons connected to it. After the input, the specific neuron integrates the combined signals and provides a non-linear response (aka. activation function or 'thresholding'). Different Non-Linear Activation Functions (NLAF) can deliver advantages in various applications, which was extensively investigated recently [11]. The implementations of those nonlinearities (neurons) have been experimentally demonstrated based on the physical representation of signals fall into two different implementations: optical-electrical-optical (O-E-O) or all-optically [12–15]. All-optical neurons can represent the signals as semiconductor carriers or optical susceptibility, but this scheme is not energy efficient. Since the optical nonlinear susceptibilities are usually very weak means it is weaker than its input and thus incapable of driving even a single other neuron; as a result, sometimes it is combined with an optical carrier regeneration scheme. An interesting and energy-efficient solution, proposed and demonstrated in this work, is to combine non-linear optical devices with optical carrier regenerations, since strong nonlinearity can be induced electro-optically. However, this approach evidently poses another challenge to the system – distinguishing between the switching light and the signal light, and an optical amplifier may need to be employed to boost the output signals, which increases the total energy consumptions of the system. Recently, all-optical non-linear modules have been experimentally or numerically demonstrated with promising results in terms of efficiency and throughput achieve by different approach such as two-section distributed-feedback lasers, induced transparency in quantum assembly, disks lasers, reverse absorption, saturable absorption or graphene excitable lasers [12–17]. Considering the complexity, a more compact and straightforward way to achieve those functionalities is to exploit electro-optic tuned non-linear material by means of an electro-optic modulators (as NFAF) connected to a photodiode (neuron signal summation) [18–20]. Compared to the other modulator approaches that required interferometric schemes, an electro-absorption modulator (EAM) has the potential to exhibit much lower conversion costs from one processing stage to another and easily be fully integrated on silicon photonic platforms. The characteristics for optimizing the EAM are the modulation bandwidth and the light-matter interaction in order to achieve high modulation speed and low energy-per-compute surpassing electronic efficiency. While having a promising performance for neural networks, the active material and device configuration need to be carefully engineered simultaneously.

The transparent conductive oxide (TCO) material class is CMOS compatible which can be easily integrated with silicon photonic platforms with on-chip electrical devices such as digital-to-analog (DAC) and analog-to-digital converters (ADC) provide the ability to make large-scale networks-on-chip. Indium Tin Oxide (ITO), a material belonging to the TCO family, has shown strong index modulation and tunable optical nonlinearity in its epsilon-near-zero frequency band. Recently, research and applications based on ITO modulators focused on energy efficiency with compact size and design, which can be densely integrated with silicon photonics platforms [21-24]. While a systematic and complete analytical and computational study for on-chip electro-optic modulators has been investigated, taking into account developing active materials, model concerns, and cavity feedback at the nanoscale, which gives a figure of merit for designing electro-optic modulators [21]. Free-carrier dispersive effects in ITO were explored for EO modulation with unity-strong index modulation. The main advantages of using ITO to traditional materials (e.g., Si) in photonic integrated circuit (PIC) are listed: first, the high carrier concentration in ITO allows higher charge modulation per unit area than Si for a given voltage. Second, because of the low permittivity of ITO, the index change per unit applied bias is significantly higher. Third, the epsilon near zero (ENZ) region can improve the modulation effects by electro-static gating. Finally, ITO has the potential to be complementary-metal-oxide-semiconductor (CMOS) compatible, as stated by major electronic manufacturers. Furthermore, Giga-Hertz-fast ITO-based plasmonic Mach-Zehnder interferometric modulator has been demonstrated by modulating the real part of the index, while at the same time the light-matter interaction enhanced by a plasmonic hybrid mode with small RC-delays [22, 23]. Also, broadband highly energy-efficient EOM has also been reported, utilizing strong index change from both real and imaginary parts of the materials, enabling compact and high-performing modulator [24].

In addition to TCOs for optical neurons are carbon-based materials such as Graphene, a single layer hexagonal lattice of carbon atoms with the massless Dirac electronic structure, which exhibits exceptional electron mobility and a constant absorption coefficient of 2.3% over a wide spectral range from visible to infrared, which makes it a promising material for many applications [25, 26]. Graphene also has shown potential for high-speed optical devices such as modulators or detectors [27, 28]. In earlier work, theoretical investigation of graphene-based EO plasmonic modulators has been reported, with high energy and modulation efficiency by integrating graphene on a dielectric-loaded surface plasmon polariton waveguides showcasing the unique properties of graphene, such as strong coupling with light, high-speed operation, and gate-dependent optical conductivity make it a promising material for the realization of novel modulators [27, 28]. Single-layer graphene can provide the highest saturable absorption for a unit of material, which enables the construction of high-efficiency electro-absorption modulators. An all-optical graphene modulator has been shown with ~2.2 ps response time by making the graphene wrapped around a single-mode microfiber, and the response time is limited only by the intrinsic carrier relaxation time of graphene. In an all-optical scheme, the carriers in the graphene are excited by the switch light. Through Pauli blocking of interband transitions, the absorption threshold shifted to a higher frequency, leading to a more negligible attenuation of the signal light. Compared to traditional silicon-based modulators, graphene device footprint, operation voltage, and modulation speed are improved significantly, owing to its strong EO properties and intrinsic carrier mobilities. Germanium and compound semiconductors face a challenge of integration with current silicon photonics platforms; however, graphene can be easily integrated with PICs without any restrictions. Waveguide-integrated graphene-based electro-absorption modulator has been reported, the modulation was achieved by actively tuning the Fermi level of a monolayer graphene sheet. The graphene-based EOM showed a high modulation efficiency over a broad range of wavelength [29]. Graphene-based Mach-Zehnder interferometer modulator also has been demonstrated with significantly large modulation efficiency of a compact design. The modulation modes can be switched between electro-absorption and electro-refractive by applying different voltages, which can improve the signal-to-noise ratio of graphene-based electro-absorption modulators in long-haul communications [30]. Furthermore, it has attracted interest for a new field of hybrid devices, such as TCOs with graphene, to improve electrical performance. At the same time, maintain the optical transmittance at a desired level. This combined film has been demonstrated as a transparent flexible hybrid electrode which shows lower sheet resistance since the carrier concentration of the surface is improved in such hybrid film materials [31]. This allows fast carrier injection and extraction from the TCO thin film to change the free carrier density, enabling rapid and robust optical modulation.

Here, we present and demonstrate introducing electro-optic nonlinearity of photonic neurons to resemble the popular ReLU activation function. We show how this nonlinearity can be engineered via a compact free carrier-based absorption modulator based on ITO/Graphene heterojunctions integrated in Silicon photonic waveguides. This EO nonlinear thresholder is a 10's of micrometer compact device with a short response delay and lower optical loss due to the optical transmittance of the ITO and graphene compared to the silicon-based modulator [32–34]. Consequently, due to the transmittance achieved in this modulator, it can be implemented in a network without extra amplifiers to further reduce required energy consumption. Providing PIC-based nonlinearity in a synergistic MAC operation-threshold scheme, is a critical path towards ensuring lower inference energy when operating P-ASICs.

## II. DEVICE DESIGN AND FABRICATION

We fabricate the neuron-thresholding modulators on an integrated Si platform with passive waveguides on a silicon-on-insulator (SOI) substrate with 220 nm epi-Si height and 500 nm width of the waveguides to facilitate 1550 nm operation (**Fig. 1**). Grating couplers optimized for TM-like mode excitation in the waveguides are used for optical I/O to the PIC. Subsequent process steps towards the active device start with depositing a small capping (passivation) oxide layer of 5 nm to isolate any parasitic resistance loads from affecting the active device and to aid the grating coupler environmental coupling efficiency. An ITO thin film of 10 nm is deposited on top of a portion of the waveguide to act as the bottom electrode of our capacitor, followed by an oxide layer to facilitate gating (**Fig. 1**). We use a moderately high-k oxide, $Al_2O_3$, for both the capping oxide and gate oxide layer of ~30 nm, and finally, place a single layer graphene sheet on top of the stack to act as the top electrode for the active capacitor and appoint relevant contact pads with Ti/Au to probe the device electrically. The active capacitor stack is fabricated using electron-beam lithography (EBL) for defining relevant patterns, ion beam deposition (IBD) for ITO deposition, electron-beam evaporation for the metals, and lift off processing. A thin ~3 nm adhesion layer of Ti is used in the Au deposition of 50 nm. IBD processing yields dense crystalline ITO films that are pinhole-free and highly uniform, and allows for a room temperature process, which does not anneal ITO (i.e. no activation of Sn carriers as to facilitate electrostatic EO tuning) [35]. Incidentally, IBD technologies are advantageous for nanophotonic device fabrication due to their precise controllability of material properties such as microstructure, non-stoichiometry, morphology, crystallinity, etc. [36, 37]. We place the top graphene on the stack by wet transfer methods and subsequent EBL patterning and plasma etching. The ~30 nm gate $Al_2O_3$ oxide layer is grown using atomic layer deposition (ALD) with a few hydrophobic surface layers to facilitate the wet transfer process. In such capacitor configuration, the carrier concentration of the ITO thin film can be altered by the potential applied, which changes the optical properties of the material leading to variations in the portion of the electric field absorbed by the thin layer. Increasing the carrier concentration level with active electrostatic gating enhances the modulation effect due to the increased free carrier absorption of the optical mode dynamics [38-40].

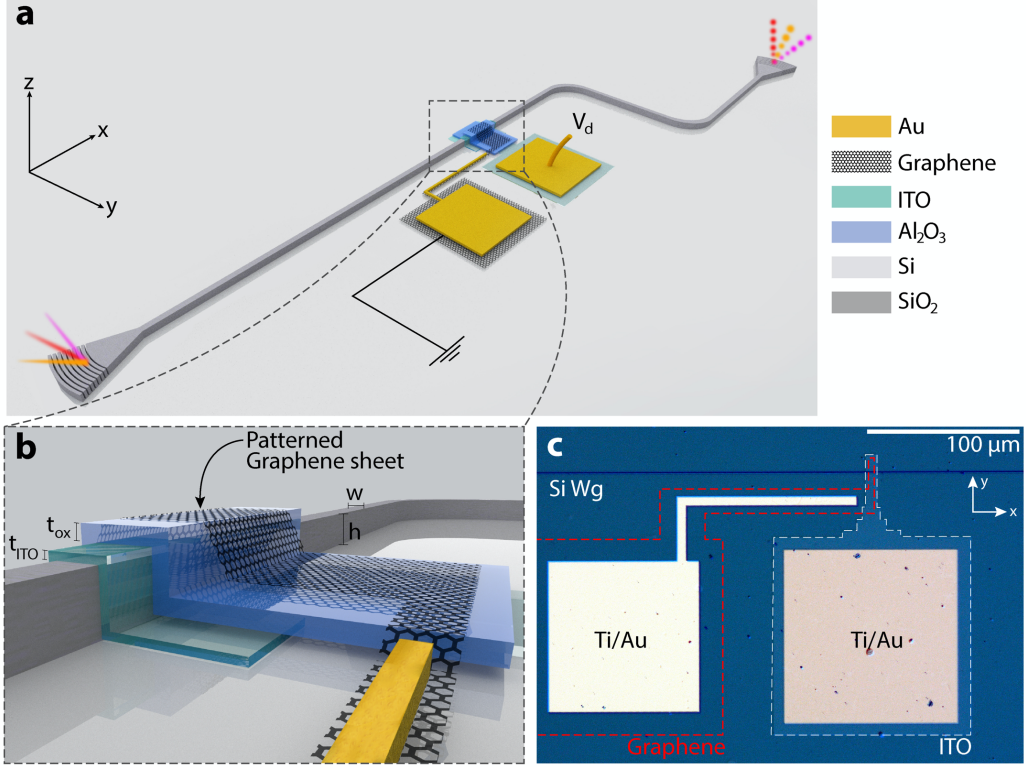

**FIG. 1 Device Details for Photonic Neuron Nonlinearity (a)** Schematic of the Graphene-ITO heterojunction absorption modulator to provide the nonlinear activation function (thresholding) to photonic neurons on the PIC. The broadband nature of absorption dynamics ensures realizing wavelength division multiplexing (WDM) operation capable of acting on multiple wavelengths simultaneously. **(b)** A closer zoomed in view of the active device region illustrating the different material layers and dimensions. Relevant parameters are: thickness of the deposited ITO thin film, $t_{ITO}$ = 10 nm; thickness of the $Al_2O_3$ gate oxide, $t_{ox}$ = 30 nm; waveguide height, $h$ = 220 nm and width, $w$ = 500 nm. Image not drawn to scale. **(c)** An optical microscope image of a fabricated device showing the active modulation region and contact pads, the dashed white and red outlines mark the patterned 10 nm ITO thin film and the patterned graphene single layer sheet on top, respectively. The overlapped region of these two (white and red) dashed outlined areas designate the active device region on top of the corresponding Si waveguide. Two contact pads side by side are deposited to facilitate biasing (marked with Ti/Au). The ITO contact is used to administer the voltage while the top graphene contact is grounded. TM-optimized grating couplers are used to couple the light from (to) the fiber into (out of) the waveguide.

A curtailing real part, $n$ and an increased imaginary part, $\kappa$ of the optical index near our operating wavelength, $\lambda$ = 1550 nm with respect to wavelength dispersion is observed in spectroscopic ellipsometry of deposited ITO thin films (see supplementary materials). This behavior is well known and expected from the Kramers-Kronig relations [38, 39]. Accumulation or depletion is obtainable through applied potential on a capacitor configuration whose one electrode is formed by ITO, changing the carrier concentration of the thin film. Note, inversion in ITO has not been reported in the literature. The optical property of ITO therefore changes dramatically depending on carrier concentration levels, resulting in strong optical modulation [21-23, 35-41]. In praxis, a 1/e decay length of about 5 nm has been measured before [42], and modulation effects have been experimentally verified over $1/e^2$ (~10 nm) thick films from the interface of the oxide and ITO [43]. The contact and sheet resistance of the ITO film is ~490 Ω and 95 Ω/□, respectively (see supplementary materials). The resistivity and mobility of the ITO film is measured to be 8.4 × 10$^{-4}$ Ω-cm and 23.7 cm$^2$/V-s, respectively; whereas the carrier concentration of the as deposited ITO film, $N_c$ = 3.1 × 10$^{20}$ cm$^{-3}$ (see supplementary materials), purposefully away from the ENZ point (~7 × 10$^{20}$ cm$^{-3}$) to keep the imaginary part low for the light-ON state of the thresholding modulator.

## III. RESULTS AND DISCUSSION

The electric field intensity $||E||^2 = |E_x|^2 + |E_y|^2 + |E_z|^2$ in the modal cross-section through the central part of the waveguide exhibits the field profile across the different layers of the active structure (Fig. 2a). The presence of significant field strength in the ITO layer compared to the monolayer graphene across the gate oxide points to the capacity of graphene in this configuration as only electrical contact refraining from any contributions to the modulation. This is also reflected by the in-plane electric-field in the mode, $|E_{in-plane}|^2 = |E_x|^2 + |E_y|^2$ (Fig. 2b) which are the field components interacting with the monolayer graphene sheet [38, 39, 44]. As the employed mode in our experiment facilitated by the grating couplers is TM-like, the in-plane electric-field is understandably inferior in strength to the overall field strength (Fig. 2b). As ITO material does not depend on the planar selectivity of the field and can alter its optical properties uniformly, the arising modulation is majorly from ITO carrier variation from accumulation/depletion based on the applied voltage because of the selected TM-like modal operation.

The device length was swept during fabrication to allow for length dependent studies of performance. We opted for a rather compact devices ranging from sub-wavelength scales, 1.4 μm to a few micron long devices, specifically 8, 12 and 15 μm. I-V measurements are performed on the active devices and all the devices show working capacitor functions in the measured voltage range (Fig. 2c). The capacitors do not exhibit hysteresis latching behavior with typical charge storage traits (see supplementary materials). In this capacitor configuration, we modulate the carrier density of the ITO thin film corresponding to the applied potential regulating the portion of the electric field absorbed by the thin layer. The mode profile showcases TM-like mode propagation in the active device with most of the light confined in the ITO (Fig. 2d). The prominent contribution in the light – matter interaction (LMI), and hence, arising modulation thereof, originates from the ITO material is also apparent from the confinement factor, Γ of the propagating mode inside the active capacitive stack. A closer investigation reveals a 2.70% Γ of the propagating mode where confinement in the ITO, $Γ_{ITO}$ is 2.66%. Only a miniscule amount of light can be felt by the graphene sheet owing to its thickness and corresponding low light confinement factor, $Γ_{graphene}$ of only 0.04%. A schematic of the modal structure is provided as a guide to the relative layers in the structure to assimilate from the modal illumination profile (Fig. 2e). A 1550 nm TM-like mode traveling in the waveguide is subjected to a shift in the mode profile due to the presence of the active capacitive stack, and accordingly, enhances the modal overlap with the active ITO layer. Increasing the carrier concentration level with active electrostatic gating enhances the modulation effect due to the increased free carrier absorption of the optical mode dynamics [38-40].

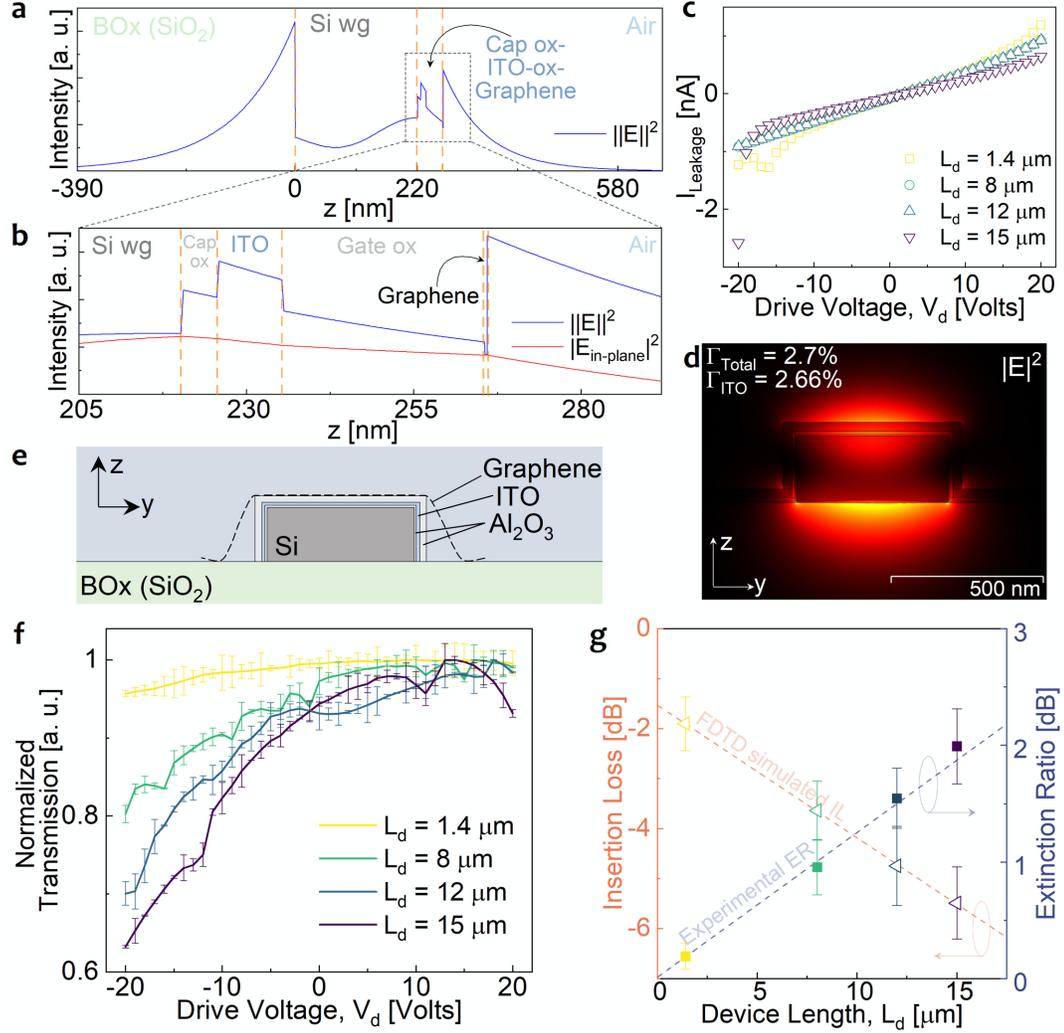

**FIG. 2 Photonic Neuron's Electro-optic Nonlinear Activation Function. (a)** Electric field intensity, $||E||^2$ along a central vertical cutline through the waveguide modulator active region showcasing different layers. **(b)** Zoom-in of (a) within the active capacitor layers, the in-plane electric field is also shown to distinguish 2-dimensoinal effects in graphene. **(c)** I–V measurements of the fabricated devices showing active capacitor functions. **(d)** FEM mode profile for the propagating TM-like mode in the active device. Most of the light is confined in the ITO. **(e)** A schematic of the modal cross-section to aid in investigating the mode profile material layers. **(f)** Normalized transmission through the fabricated devices with varying active lengths responsive to applied drive voltages, $V_d$ (Volts). **(g)** Performance metrics of the fabricated modulators including insertion loss, IL (dB, left vertical-axis, triangles) and extinction ratio, ER (dB, right vertical-axis, squares) for variations of the active modulator length, $L_d$ (μm) among fabricated devices. The filled symbols represent experimentally measured data and the empty symbols represent simulated results.

Experimental results show a modulation depth (i.e. ER) of ~2 dB for an absorption modulator length of only 15 μm (Fig. 2f) which is obviously higher per unit length compared to Si while both (ITO and Si) operate with the free carrier modulation mechanism. This improvement of ITO can be attributed to: (a) 2-3 orders higher carrier density, and (b) the higher bandgap, which consequently leads to a lower refractive index [21-23, 35, 38-41]. If the change of the carrier concentration $\delta N_c$ (e.g. due to an applied voltage bias) causes a change in the relative permittivity (dielectric constant) $\delta\varepsilon$, the corresponding change in the refractive index can be written as $\delta n = \delta\varepsilon^{1/2} \sim \delta\varepsilon/2\varepsilon^{1/2}$; hence, the refractive index change is greatly enhanced when the permittivity, $\varepsilon$ is small [21-23, 35, 38-41]. The Pauli blocking effect from graphene can be seen in the forward voltage range slightly rectifying the modulator behavior owing to the small light confinement in graphene. The extinction ratio and insertion loss both show linear trends for device length variations, as expected (Fig. 2g). The extinction ratio increases monotonically with 0.132 dB/μm slope featuring the performance for this structure within the applied voltage range with a coefficient of determination for the linear fit, $R^2 = 0.99$. Insertion loss, on the other hand, features monotonic increase with some added length independent losses from other sources, such as the grating couplers, coupling to (and from) the Si waveguides, waveguide bending losses, measurement factors, etc. Experimental results showed an insertion loss of about 0.31 dB/μm for the fabricated devices with a coefficient of determination for the linear fit, $R^2 = 0.86$. The length independent losses in our experiments amount to be almost 60 dB. This is quite high compared to our previous results [22, 23], and from experience working with similarly processed foundry tape outs. We have found the length independent losses to be around 20-30 dB in similar previous experiments. The additional almost 30-40 dB of loss can be accounted to the imperfections in our wet etch process during fabrication of this structure leading to compromised performances. We needed to fashion openings on the gate oxide film on top of the ITO contact pads to facilitate electrostatic gating, and the wet etch process timings were misjudged and many of the opening patterns were undercut by the wet etchant forcing the graphene adhesion to the top hydrophobic layer of the oxide to become loose and contaminate the entire chip. This unintended phenomenon coupled with wet transferred graphene causing wrinkling effect in some places due to the uneven patterned surface and multilayer formation in patches which could not be plasma etched with recipes aimed at monolayers affected the loss drastically. However, the low loss per length suggests feasibility of realizing longer devices to achieve higher modulation depths and can benefit in availing high speed alternatives in photonic devices refraining from plasmonic routes. To approximate for the length independent coupling losses, i.e. to and from the underlying Si waveguide, we performed FDTD simulations and found that the length dependent insertion loss matches nearly well with our experimental results at 0.27 dB/μm (Fig. 2g). FDTD results point to a coupling loss of only 0.76 dB/coupling facet, which is indicative of the feasibility of such device configurations while staying in the photonic domain without having to opt for selective doping the Si waveguide but still availing pathways for high speed operations by utilizing high-mobility single layer graphene. This can certainly pose as a promising alternative to plasmonics keeping the insertion losses minimal.

The demonstrated EAMs exhibit a modest modulation range in rather compact (linear) footprints (∼0.13 dB/μm) and considerably low insertion losses (<2 dB). Therefore, this ITO-graphene hybrid approach exemplifies a practical alternative to other EO modulators (e.g. Si and LiNbO$_3$) [21-23, 40], without necessitating any interferometric or cavity schemes and therefore, characterized by a broadband response. Next, we implement an optical module of a neuromorphic activation function based on our experimental EAM results for weighting scheme relying on WDM as EAMs are spectrally broadband by definition (no resonance used).

## III. BROADCAST AND WEIGHT PHOTONIC NEURAL NETWORK

A broadcast and weight photonic neural network [45] assigns each neuron a dedicated wavelength and multiplexing all of the outputs onto a single bus between each layer (Fig. 3). Individual nodes (Fig.3a) connect to the input bus, each receiving the output of all the nodes of the previous layer. Each color is separated from the bus and weighted by detuning a ring modulator. The output of the weight is summed onto one or more photodiodes to generate an electrical output of the accumulated signal. The electrical output is then amplified and remodulates an optical signal using an absorption modulator. The combined electrical and photonic response of the photodiode, amplifier, and modulator creates a nonlinear activation function for the neurons MAC signal. The degree of nonlinearity in neural network nodes has been shown to be important in the higher layers of a deep neural network [46] where the model must accentuate information from helpful dimensions while eliminating unhelpful dimensions to avoid the problem of dimensionality when making a decision.

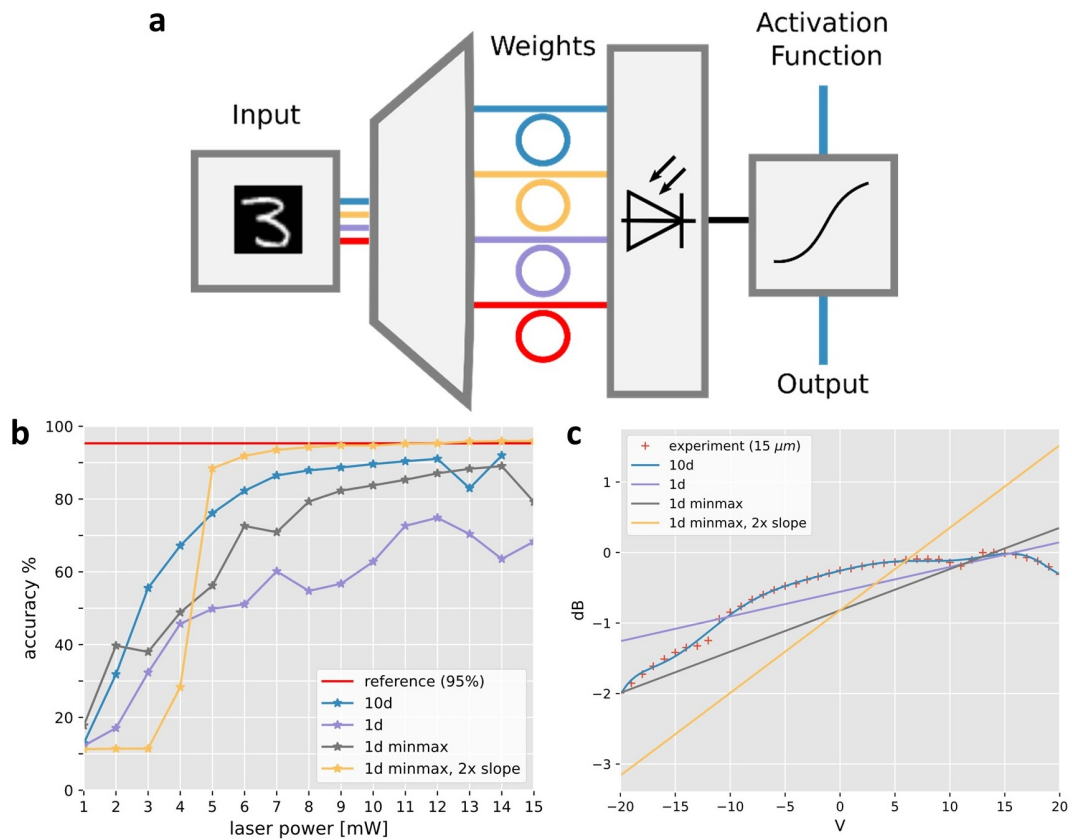

**FIG. 3 Neuromorphic nonlinear activation implementation and MNIST simulation results. (a)** Each optical neural network node demultiplexes a WDM input signal from the previous layer or input data, applies weights with ring modulators to each input color, integrates the input signals with a photodiode and modulates the output onto a new optical signal with an optical modulator. **(b)** MNIST simulation results using activation functions fit to the transmission transfer function of the 15 μm modulator **(c)**, shows accuracy increasing with transfer function steepness. While the nonlinear 10-dimensional polynomial fit (10d) shows increased accuracy when compared with the linear least squares fit (1d) and the linear fit through the minimum and maximum point (1d minmax), increasing the slope by 2x (1d minmax, 2x slope), generates the highest accuracy.

The 15 μm modulator transmission transfer function was fit to a nonlinear 10-dimensional polynomial, a linear least square fit, a line passing through the minimum and maximum point, and a hypothetical line with twice the slope as the line through the minimum and maximum points (Fig. 3c). These activation functions were, exemplary, inserted into a Modified National Institute of Standards and Technology (MNIST) database [47] neural network model consisting of three layers: two 100 node, fully-connected optical layers using activation functions using the fit transfer functions, followed by a simulated electronic dense 10 node softmax activation output layer. The models were trained in Python with Keras using the Adagrad method with a 0.005 learning rate for 1000 epochs with a 1024 batch size. The optical-link noise model assumed an operating frequency of 1 GHz, 300 K temperature, 40 dB gain between layers, 50 Ω photodiode impedance, 5 fF photodiode capacitance, and 50 pA photodiode dark current.

The simulation results show close to 10% increased accuracy for the nonlinear 10 dimensional fit when compared to both the least squares linear fit and the linear fit through the minimum and maximum points. To determine if this increased performance was due to the nonlinear shape of the transfer function or simply to the greater slope in some regions of the nonlinear transfer function, a hypothetical transfer function was modeled based on the line through the minimum and maximum points but with twice the slope. This hypothetical linear transfer function outperformed the other models at laser powers above 4 mW. While this test demonstrates that modulators with linear transfer functions may outperform nonlinear transfer functions in some cases, it is physically more challenging with a greater modulation depth and gain above ~5 V.

## V. CONCLUSION

In this work, we demonstrated the first ITO-graphene based heterojunction scheme for an electro-absorption modulator to perform an electro-optic nonlinearity (thresholding) on photonic neurons inside PIC-based neural networks. These electro-optic thresholders integrated into a Silicon photonic platform enable monolithic nonlinearity without the need for dual chip approaches which is costly from an energy-driver link budget aspect (50pJ/bit for off-chip versus ~1pJ/bit for on-chip signal routing). These ITO-graphene photonic heterojunctions enable realizing ReLU-like transfer function for thresholding in photonic ASICs. Our results show a path for alternatives to plasmonic solutions keeping the insertion losses low availing a photonic paradigm for high-speed operations utilizing ITO processing synergies. We further showed neuromorphic application feasibility of the demonstrated absorption modulators with the same enabling nonlinear activation functionalities in a feed forward broadcast and weight photonic neural network benchmarked by means of the MNIST classifier.

## SUPPLEMENTARY MATERIAL

See supplementary material for additional information.

## ACKNOWLEDGMENTS

V.S. is supported by the Air Force Office of Scientific Research (FA9550-20-1-0193) under the PECASE Award.

## AVAILABILITY OF DATA

The data that supports the findings of this study are available within the article [and its supplementary material].

## AUTHOR CONTRIBUTIONS

R.A., Z.M. and V.S. initiated the project and conceived the experiments. R.A. designed and fabricated the devices. R.A. and R.M. conducted experimental measurements. R.A. and Z.M. performed supporting experiments. R.A. and H.W. conducted simulations and data analysis. J.K.G. conducted the neural network simulations. H.D. and J.B.K. provided suggestions throughout the project. V.S. supervised the project. All authors discussed the results and commented on the manuscript.